# Low-bias Negative Differential Resistance effect in armchair graphene nanoribbon junctions

*Suchun Li[1,2,3], Chee Kwan Gan[2], Young-Woo Son[4], Yuan Ping Feng[1,3], Su Ying Quek[1,2,a)]*

[1] Department of Physics, Faculty of Science, National University of Singapore, 2 Science Drive 3, Singapore 117551

[2] Institute of High Performance Computing, Agency for Science, Technology and Research, 1 Fusionopolis Way, #16-16 Connexis, Singapore 138632

[3] NUS Graduate School for Integrative Sciences and Engineering, National University of Singapore, 28 Medical Drive, Singapore 117456

[4] Korea Institute for Advanced Study, Seoul 130-722, Korea

[a)]Corresponding author    Tel: (65) 6601 3640   Fax: (65) 6777 6126   Email: phyqsy@nus.edu.sg

**Abstract**

**Graphene nanoribbons with armchair edges (AGNRs) have bandgaps that can be flexibly tuned via the ribbon width. A junction made of a narrower AGNR sandwiched between two wider AGNR leads was recently reported to possess two perfect transmission channels close to the Fermi level. Here, we report that by using a bias voltage to drive these transmission channels into the gap of the wider AGNR lead, we can obtain a negative differential resistance (NDR) effect. Owing to the intrinsic properties of the AGNR junctions, the on-set bias reaches as low as ~ 0.2 V and the valley current almost vanishes. We further show that such NDR effect is robust against details of the atomic structure of the junction, substrate and whether the junction is made by etching or by hydrogenation.**

Reducing two-dimensional graphene into quasi one-dimensional graphene nanoribbons with armchair edges (AGNRs) is an effective way to open a bandgap from the quantum confinement effect. Both theoretical (zone-folding approximations[1], orbital tight-binding models[2,3], and first principles calculations[4-6]) and experimental studies predict a gap $E_g$ scaling inversely with the AGNR width. This semiconducting behavior overcomes the gapless limitation in graphene and leads to many promising electronic applications, such as graphene transistors[7-9] and tunnel barriers. On the experimental side, tremendous advancements have been made recently in bottom-up chemical synthesis of AGNRs with controlled widths[10-14], nanostructuring of graphene[15-19], and formation of AGNR heterojunctions[20]. It is therefore important to explore the potential of such well-defined semiconducting AGNRs for applications in nanoelectronics, especially all-carbon devices where all discrete electrical components are patterned out of the same piece of graphene sheet and seamlessly connected to one another.

Recently, an all-carbon junction made of a narrower AGNR connected to two wider AGNR leads was reported to posses two intrinsic transmission channels close to the Fermi level $E_F$ [21]. In this letter, we show that by utilizing these transmission channels and the gap of the semiconducting AGNR leads, we can obtain a negative differential resistance (NDR) effect. The NDR effect is generated via the resonant tunneling model, as first proposed by Rakshit *et al.* for the system of a molecule adsorbed on a silicon surface [22]. Owing to the intrinsic properties of these transmission channels, the on-set bias of this NDR effects is only on order of 0.1 V, which is one order of magnitude lower than in the molecule-on-silicon system (above 2 V even for the shortest tip-sample distance) [23].

Following previous conventions [5], we refer to an AGNR with *n* carbon atoms spanning its width (see Fig. 1a) as an *n*-AGNR. Without loss of generality, we choose a prototypical AGNR junction made of a 5-AGNR segment connected seamlessly to 17-AGNR leads on both sides (see Fig. 1a) as an example to demonstrate the mechanism of NDR. We refer to this structure as a 17-5-17-AGNR junction. The 17-AGNR lead is semiconducting with a bandgap ~ 0.1eV [5]. To make it metallic, the 17-AGNR lead region is doped with 0.15% of boron atoms via the virtual crystal approximation [24]. This doping shifts the Fermi level from the middle of the gap to ~0.05eV below the valence band edge, as indicated in Fig. 1b.

First principles transport calculations are performed using the non-equilibrium Green function (NEGF) as implemented in TranSIESTA [25]. We use the local density approximation (LDA) for the exchange-correlation functional, as implemented in SIESTA [26], and a double-zeta basis-set.

Each carbon atom at the edges is passivated with a hydrogen atom and the structure is fully relaxed until the forces on atoms are less than 0.01 eV/Å.

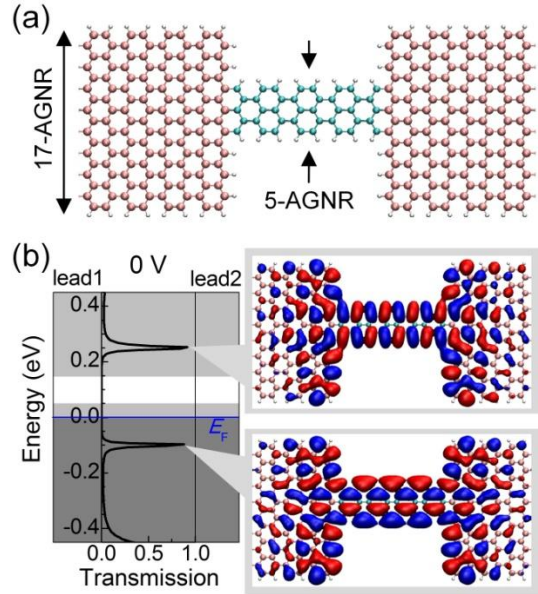

Fig. 1 (a) Atomic structure of a 17-5-17-AGNR junction. The 17-AGNR regions are doped with 0.15% of boron atoms via the virtual crystal approximation[24]. The blue, brown, and white balls denote pure carbon atoms, virtual atoms composed of 99.85% of carbon plus 0.15% of boron, and hydrogen atoms, respectively. (b) Transmission curve for the junction shown in (a) at zero bias. The zero energy is set at the Fermi level $E_F$, as indicated by the blue line. The dark and light gray shadings (below and above $E_F$) indicate energy ranges with occupied and unoccupied states in the leads, respectively. The white blank region indicates the gap of leads. Inset: the real parts of eigenchannel wavefunctions (isovalue = +/- 0.025) at the two transmission peaks. The imaginary parts show similar features.

This 17-5-17-AGNR junction shows two narrow transmission peaks near $E_F$ (Fig. 1b), consistent with predictions in Ref. [21]. According to their eigenchannel wavefunction patterns (Fig. 1b inset), we can identify that these two transmission peaks are the same as those studied in Ref. [21], that is, they are bonding and antibonding couplings of states related to the zigzag edge and the 5-AGNR (see Ref. [21] for details). Based on this fact, we refer to the peak below $E_F$ as the bonding peak, and the one above $E_F$ as antibonding peak. As will be demonstrated in detail in the next section, we can use a small bias voltage to drive these peaks into the gap of leads to obtain NDR.

Figure 2a gives the IV curve of the 17-5-17-AGNR junction, where an NDR effect with an on-set bias of only 0.2V and a peak-to-valley current ratio > 6 is observed. This on-set bias is one order of magnitude lower than that observed in the molecule-on-silicon system (above 2 V even for the shortest tip-sample distance)[23]. The peak-to-valley ratio is also higher than that observed in most other NDR systems, such as the GNR-CNT cross bar system with a peak-to-valley ratio of 2 [27] and the Si-SiGe system with a peak-to-valley ratio of 5.2 [28].

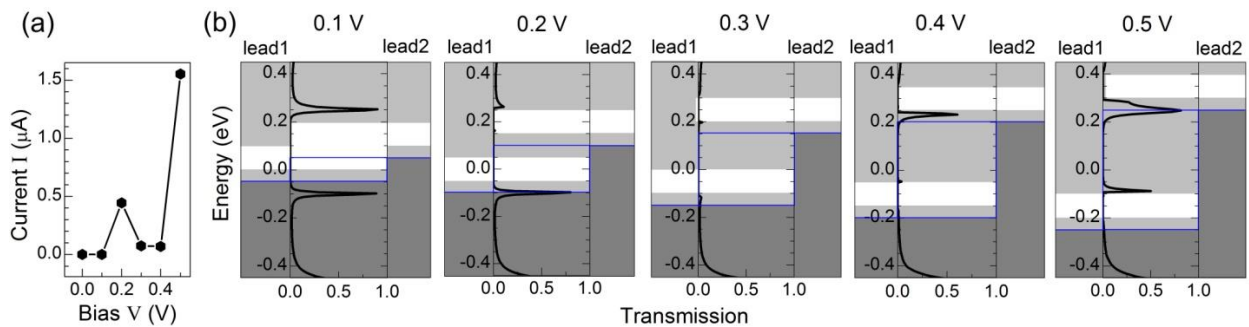

Fig. 2 (a) IV curve of the 17-5-17-AGNR junction shown in Fig. 1 (a). (b) Transmission spectra of the junction with a bias of 0.1 to 0.5V. The dark and light gray shadings indicate energy ranges with occupied and unoccupied states in the leads, respectively. The white blank regions indicate the gaps of leads. The blue box indicates the biased energy window, within which one lead is occupied and the other is unoccupied.

The origin of NDR can be understood by examining how the transmission curve varies as a function of the bias voltage, as shown in Fig. 2b. When we apply a bias of 0.1 V, the Fermi level of lead 1 ($E_{F1}$) drops by 0.05 eV and the Fermi level of lead 2 ($E_{F2}$) increases by 0.05 eV. This produces an energy window of 0.1 eV, as indicated by the blue box (Fig. 2b, 0.1 V). Within this energy window, states in lead 2 are occupied while states in lead 1 are empty at low temperatures. The low temperature current is the integration of the transmission probability $T(E)$ within this energy window $I = \frac{2e}{h} \int_{E_{F1}}^{E_{F2}} T(E) dE$ [29]. Since the cumulative transmission probability falling into the energy window at the bias of 0.1 V is almost zero, the current is also almost zero.

As the bias increases to 0.2 V, the bias window widens and partially covers the bonding transmission peak, which gives rise to a current up to ~ 0.444 µA. At the bias of 0.3 V and 0.4 V, although the bonding channel would be well within the energy window, it is however driven into the gap of lead 1, where there are no incoming electrons, hence giving no transmission. This leads to a significant drop of the current from ~ 0.444 µA at 0.2 V to only ~ 0.071 µA at 0.4 V, thus results in a negative difference resistance with a peak-to-valley ratio ~ 0.444/0.071 = 6.25. For a typical junction without such a gap in the lead, the current usually increases monotonically with the bias, as a larger bias window covers more transmission probabilities.

At a bias of 0.5V, as the gap of lead 1 shifts further down, the bonding peak partially emerges out of the gap. Moreover, the antibonding transmission peak also starts to fall into the biased energy window, so together the two transmission peaks give rise to a current as large as ~ 1.6 µA. For bias beyond 0.5 V, both transmission peaks will fall into the biased energy window.

Moreover, these peaks will no longer be affected by the gaps of leads that are moving away. Therefore, we expect the current to increase further.

We see from the mechanism discussed above that the AGNR junction possesses many advantages to achieve NDR effect with low on-set bias and high peak-to-valley ratio. Firstly, the bonding and antibonding nature of the two transmission peaks determines that they are close to $E_F$. For example, the bonding peak of the 17-5-17-AGNR junction is only ~0.1 eV below $E_F$ (Fig.1b). The closeness to $E_F$ makes it easier to achieve NDR with a relatively low on-set bias.

Secondly, the eigenchannel wavefunctions (Fig. 1b) of the transmission peaks also show that the two states are strongly localized over the middle AGNR region and the zigzag edges at the two interfaces, but relatively weakly coupled to the leads. The localization is an intrinsic property originated from the zigzag edge state[21]. On one hand, the weak coupling determines that the states are not perturbed much by the bias applied to the two leads. Figure 2b shows that the energy and shape of these transmission peaks do not vary much as bias increases (except a suppression by the gap). In addition, the eigenchannel wavefunction at the bonding peak under a bias of as large as 0.5 V (see Fig. 3) remains essentially the same as that at 0 V (Fig. 1b inset), further confirming that the state is almost not perturbed by the bias. It would be more difficult or even impossible to achieve NDR if these states were drifted away or destroyed by the bias. On the other hand, the weak coupling determines that the two transmission peaks are very narrow (energy width < 0.05 eV), thus enabling the peaks to be completely suppressed by the gap of the lead (~0.1 eV for 17-AGNR, see Fig. 1b and Fig. 2b)). In addition, there is no other transmission peaks nearby. Therefore, once the bonding peak is suppressed by the gap of the lead, we can achieve an almost *zero valley current*.

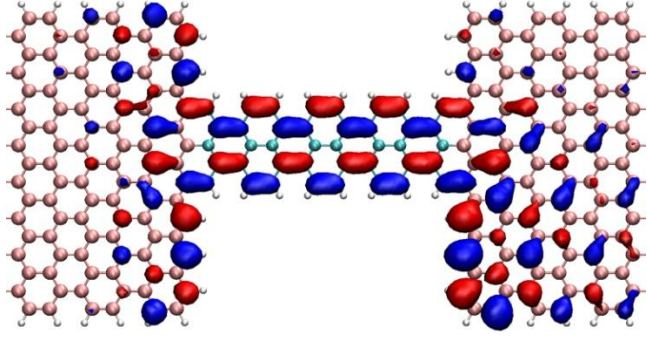

Fig. 3. Real part of the eigenchannel wavefunction with isovalue = +/- 0.025 at the bonding peak in Fig. 2 (b) under a bias of 0.5 V. The imaginary part shows the same feature.

Thirdly, the two narrow transmission peaks close to $E_F$ are intrinsic to the AGNR junction as long as the middle AGNR is in the $3p+2$ family and both interfaces are terminated with zigzag edges[21]. The gap in the AGNR lead is also intrinsic and can be well controlled through the width. Therefore, the high performance NDR that we demonstrated using the example of 17-5-17-AGNR can be generalized to other AGNR junctions.

Fourthly, these intrinsic channels still persist when the junction is put on graphene or boron-nitride substrates [Supplementary Fig. S1]. This could be understood since the two channels are originated from the π-electrons of carbon atoms[21], which do not interact much with these substrates. Therefore, we expect the high performance NDR would also be robust against substrates. This provides great convenience in practical applications.

So far, we have demonstrated NDR in an etched AGNR junction. As the perfect transmission channels giving rise to NDR originate from the π-electrons of carbon atoms[21], we expect that an equivalent junction of π-electrons (rather than carbon atoms) is enough to exhibit the bonding and antibonding states and the relevant NDR effects. Figure 4a shows one such junction of π-electrons made from perfect 17-AGNR by passivating the π-electron of extra carbon atoms with a hydrogen atom; we call this a hydrogenated junction. This hydrogenated junction shows essentially the same transmission curve at zero bias (Fig. 4c) as that of an etched junction (Fig. 1b). The eigenchannel wavefunctions at the two narrow peaks close to $E_F$ (Fig. 4d) look similar as the eigenchannels determining the NDR in an etched junction (Fig. 1b), confirming that the nature of the channels are the same. As a result, this hydrogenated junction gives an IV curve (Fig. 4b) also similar to that of the equivalent etched junction, which exhibits an NDR with the same on-set voltage of 0.2 V and an even larger peak-to-valley ratio of ~18.6.

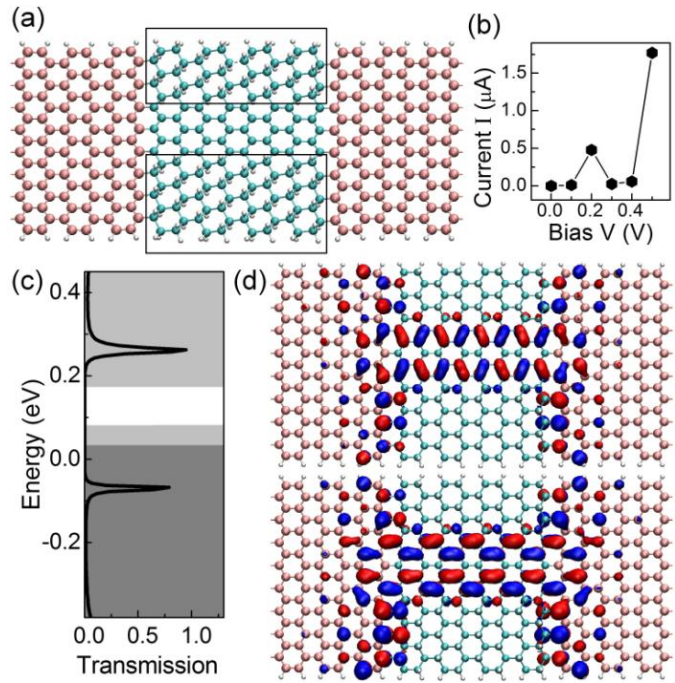

Fig. 4 The (a) geometry structure, (b) IV curve, and (c) transmission curve at 0V of the 17-5-17-AGNR junction made from 17-AGNR by hydrogenation. The boxes in (a) indicate the extra carbon atoms, and each of them is passivated with an additional hydrogen atom. (d) The real

parts of eigenchannel wavefunctions at the two perfect transmission peaks with isovalue = +/- 0.08. The imaginary parts show similar features.

In conclusion, we have demonstrated negative differential resistance in AGNR junctions by utilizing intrinsic transmission channels of the junction and the gap of the semiconducting AGNR leads. These intrinsic channels are close to $E_F$, weakly coupled to the leads, and there are no other channels nearby. All these properties help to give a very low on-set voltage and high peak-to-valley ratio for the NDR effect. In addition, the intrinsic channels and hence the resulting NDR effect are robust against effects from substrates, with the junction made by either etching or hydrogenation. Given recent experimental advancements in fabrication of AGNRs and related nanostructures, we expect this robust high performance NDR effect in AGNR junctions could lead to useful applications in graphene based all-carbon nanoelectronics.

**Acknowledgements**

S.L. and S.Y.Q. thank A*STAR for funding through the A*GS Scholarship and IHPC Independent Investigatorship respectively. S.Y.Q. also thanks the Singapore NRF for funding through the NRF Fellowship (NRF-NRFF2013-07). We thank A*CRC for computational support, and CH Sow, X Luo and N Gorjizadeh for discussions. Y.-W. S. was in part supported by the NRF grant funded by Korea MEST (QMMRC, R11-2008-053-01002-0 and Nano R&D program 2008-03670).